\documentclass[prl,twocolumn,aps,showpacs,preprintnumbers,amsmath,amssymb, superscriptaddress]{revtex4-1}
\pdfoutput=1

\usepackage{bm}
\usepackage{graphicx}
\usepackage{color}
\usepackage{SIunits}

\begin{document}

\title{Angle-resolved spectroscopy study of Ni-based superconductor SrNi$_2$As$_2$}

\author{L.-K. Zeng}
\affiliation{Beijing National Laboratory for Condensed Matter Physics, and Institute of Physics, Chinese Academy of Sciences, Beijing 100190, China}

\author{P. Richard}\email{p.richard@iphy.ac.cn}
\affiliation{Beijing National Laboratory for Condensed Matter Physics, and Institute of Physics, Chinese Academy of Sciences, Beijing 100190, China}
\affiliation{Collaborative Innovation Center of Quantum Matter, Beijing, China}

\author{A. van Roekeghem}
\affiliation{Beijing National Laboratory for Condensed Matter Physics, and Institute of Physics, Chinese Academy of Sciences, Beijing 100190, China}
\affiliation{Centre de Physique Th{\'e}orique, Ecole Polytechnique, CNRS-UMR 7644, 91128 Palaiseau, France}

\author{J.-X. Yin}
\affiliation{Beijing National Laboratory for Condensed Matter Physics, and Institute of Physics, Chinese Academy of Sciences, Beijing 100190, China}

\author{S.-F. Wu}
\affiliation{Beijing National Laboratory for Condensed Matter Physics, and Institute of Physics, Chinese Academy of Sciences, Beijing 100190, China}

\author{Z. G. Chen}
\affiliation{Beijing National Laboratory for Condensed Matter Physics, and Institute of Physics, Chinese Academy of Sciences, Beijing 100190, China}

\author{N. L. Wang}
\affiliation{International Center for Quantum Materials, School of Physics, Peking University, Beijing 100871, China}
\affiliation{Collaborative Innovation Center of Quantum Matter, Beijing, China}

\author{S. Biermann}
\affiliation{Centre de Physique Th{\'e}orique, Ecole Polytechnique, CNRS-UMR 7644, 91128 Palaiseau, France}
\affiliation{Coll\`{e}ge de France, 11 place Marcelin Berthelot, 75005 Paris, France}
\affiliation{European Theoretical Synchrotron Facility, Europe}

\author{T. Qian}\email{tqian@iphy.ac.cn}
\affiliation{Beijing National Laboratory for Condensed Matter Physics, and Institute of Physics, Chinese Academy of Sciences, Beijing 100190, China}
\affiliation{Collaborative Innovation Center of Quantum Matter, Beijing, China}

\author{H. Ding}\email{dingh@iphy.ac.cn}
\affiliation{Beijing National Laboratory for Condensed Matter Physics, and Institute of Physics, Chinese Academy of Sciences, Beijing 100190, China}
\affiliation{Collaborative Innovation Center of Quantum Matter, Beijing, China}

\date{\today}
\begin{abstract}
We performed an angle-resolved photoemission spectroscopy study of the Ni-based superconductor SrNi$_2$As$_2$. Electron and hole Fermi surface pockets are observed, but their different shapes and sizes lead to very poor nesting conditions. The experimental electronic band structure of SrNi$_2$As$_2$ is in good agreement with first-principles calculations after a slight renormalization (by a factor 1.1), confirming the picture of Hund's exchange-dominated electronic correlations decreasing with increasing filling of the $3d$ shell in the Fe-, Co- and Ni-based compounds. These findings emphasize the importance of Hund's coupling and $3d$-orbital filling as key tuning parameters of electronic correlations in transition metal pnictides. 
\end{abstract}


\maketitle
\section{INTRODUCTION}
Although isostructural counterparts of the Fe-based superconductors La(O,F)FeAs \cite{Kamihara_JACS2008} and (Ba,K)Fe$_2$As$_2$ \cite{Rotter_PRL101} can be synthesized using $3d$ transition metal atoms other than Fe, superconductivity is found only in Ni-based LaONiPn \cite{LaNiOP,Z_Li_PRB78} and ANi$_2$Pn$_2$ \cite{PhysRevB.79.132506,0953-8984-20-34-342203,PhysRevB.78.172504} (A = Ca, Sr and Ba, Pn = P and As), albeit with much lower superconducting transition temperature ($T_c$) values than in the Fe-based superconductors. Unlike the Fe-based family of superconductors, the parent compounds of the Ni-based superconductors do not show any magnetic ordering. While electronic correlations are responsible for a renormalization of the band structure of the Fe-based superconductors by a typical factor of 2$\sim$5 \cite{0034-4885-74-12-124512, AmbroiseCRPhys2016}, recent angle-resolved photoemission spectroscopy study (ARPES) reported that the band dispersions are renormalized by a factor of 1$\sim$2 in BaNi$_2$P$_2$ \cite{PhysRevB.89.195138} and by a factor of 1.4 in TlNi$_2$Se$_2$ \cite{N_Xu_PRB92}, suggesting weaker electronic correlations in the Ni-based superconductors. Since Hund's coupling is believed to play a significant role to tune the correlation strength in these materials, a systematic study and comparison is of importance. Investigating the details of the electronic structure of the 122-NiPn materials thus appears a good starting point for understanding the key parameters for Fe-based superconductivity. Among the Ni-based superconductors, SrNi$_2$As$_2$  is particularly interesting. Superconductivity in this material is found at $T_c = 0.62$ K, and unlike BaNi$_2$As$_2$ and SrNi$_2$P$_2$, SrNi$_2$As$_2$ share the same tetragonal crystal structure as (Ba,K)Fe$_2$As$_2$ at high temperature, and no structural transition was reported in SrNi$_2$As$_2$ upon cooling. 


In this paper, we present ARPES results on the Ni-based superconductor SrNi$_2$As$_2$. We observe small electronic correlations as compared with the Fe-based superconductors, without detectable orbital-selectivity. We show that although the Hund's coupling tunes the correlations in the Fe- compounds, its effect is limited in the Co- and Ni- compounds due to the large $3d$ shell electron filling. Therefore, in $3d$ compounds both Hund's coupling and shell electron filling play important roles in explaining the electronic correlations. The absence of magnetic order in SrNi$_2$As$_2$ is probably associated with the poor Fermi surface nesting conditions derived from the measured Fermi surfaces, as well as the negligible local moment due to the weakness of the electronic correlations. 

\begin{figure}[!t]
\begin{center}
\includegraphics[width=\columnwidth]{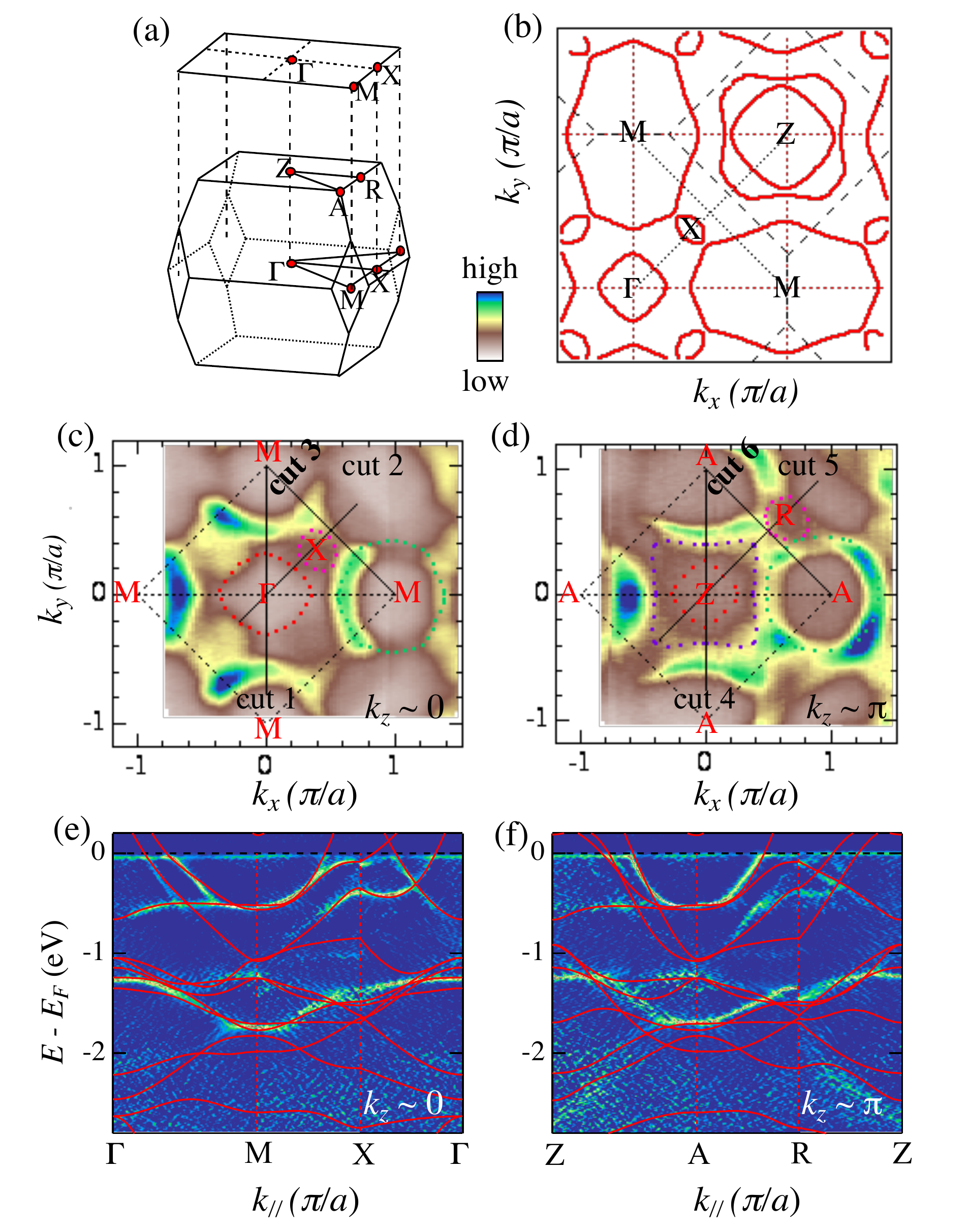}
\end{center}
\caption{\label{fig1_map_GMX}(Color online) (a) 3D BZ and projection in the 2D BZ, along with the definitions of high-symmetry points. (b) Calculated LDA FS. (c) and (d) ARPES mapping intensity plots at $E_F$ recorded at 30 K on SrNi$_2$As$_2$ in the $k_z\sim 0$ and $k_z\sim\pi$ planes, respectively. The intensity is obtained by integrating the spectra within $\pm$5 meV with respect to $E_F$. The $k_F$ positions are indicated by dots. (e) and (f) 2D curvature intensity plot along high-symmetry lines recorded at 30 K in the $k_z\sim 0$ and $k_z\sim\pi$ planes, respectively. The LDA band structure renormalized by a factor 1.1 is overlapped for comparison.}
\end{figure}

\section{\label{sec:level2}EXPERIMENT}
High-quality single crystals of SrNi$_2$As$_2$ were grown by the flux method. ARPES experiments were performed at beamline PGM of the Synchrotron Radiation Center (Wisconsin) equipped with a Scienta R4000 analyzer. The energy and angular resolutions were set at 15-30 meV and 0.2\degree, respectively. The samples were cleaved \emph{in situ} and measured in a vacuum better than 5$\times$10$^{-11}$ Torr. The Fermi level ($E_F$) of the samples was referenced to that of a gold film evaporated onto the sample holder. The high-symmetry points associated to the first Brillouin zone (BZ) of SrNi$_2$As$_2$ are defined in Fig. \ref{fig1_map_GMX}(a), along with their two-dimensional (2D) projections. Following our previous notations, the various cuts are indexed using the 1 Ni/unit cell, with the parameter $a$ corresponding to the distance between two Ni atoms.

\section{\label{Results}RESULTS  AND DiSCUSSION}

\begin{figure*}[!t]
\begin{center}
\includegraphics[width=7.2in]{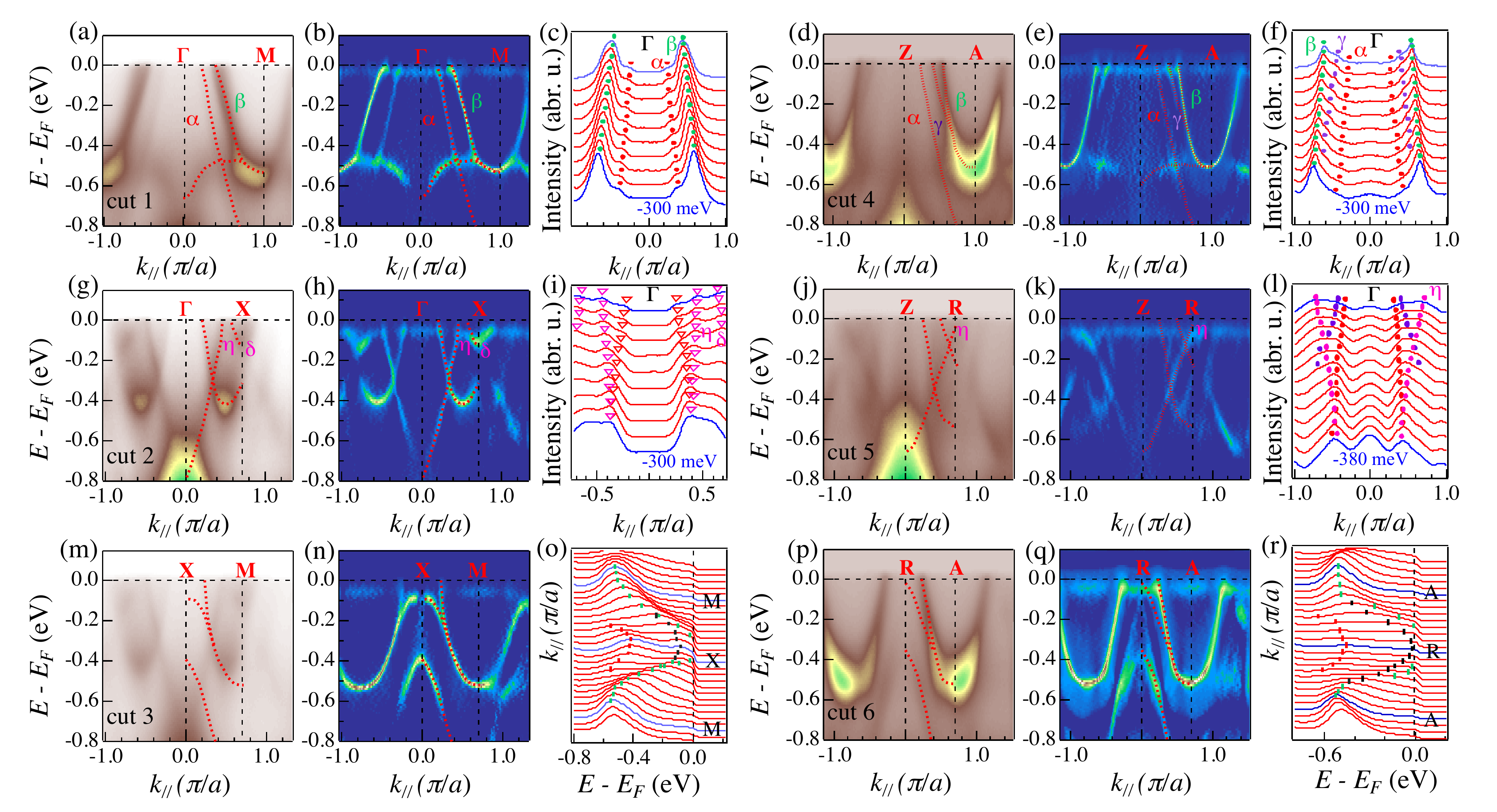}
\end{center}
\caption{\label{fig2_Ascuts}(Color online) ARPES data recorded on SrNi$_2$As$_2$ along different high-symmetry lines (the cuts correspond to the ones in \ref{fig1_map_GMX}(c) and (d)). The 3 left columns refer to $k_z=0$ while the 3 right columns correspond to $k_z=\pi$. The first (a, g, m) and fourth (d, j, p) columns correspond to ARPES intensity plots. The second (b, h, n) and fifth (e, k, q) columns correspond to the intensity of 2D curvature of the corresponding ARPES intensity plots. The third and sixth columns are associated to either MDC (c, i, f, l) plots or EDC plots (o, r) of the corresponding ARPES intensity plots. The peak positions are identified with symbols.}
\end{figure*}

The Fermi surface (FS) mappings of Sr$_2$Ni$_2$As$_2$ obtained by integrating the ARPES intensity within $\pm 5$ meV of $E_F$ in the  $k_z\sim 0$ and $k_z\sim\pi$ planes are displayed in Figs. \ref{fig1_map_GMX}(c) and \ref{fig1_map_GMX}(d), respectively. The Fermi wave vector ($k_F$) positions extracted from the band dispersions are also plotted. The results are quite different from those obtained for the Fe-based superconductors \cite{0034-4885-74-12-124512}. At $k_z\sim 0$, a rather squarish, relatively small hole pocket is detected at the $\Gamma$ point, and a large electron pocket with an approximately hexagonal shape is observed at the M point. In addition, a small hole pocket is found slightly away from the X point. This FS topology is very similar to that measured in the $k_z\sim\pi$ plane, except that an additional hole pocket is found at Z for $k_z\sim\pi$, suggesting that one band has a strong 3D character while the other FSs are more quasi-2D. We also note that the sizes and shapes of the $\Gamma$-centered hole pockets differ significantly from those of the M-centered electron pocket, which means that the electron-hole quasi-nesting is poor in this system. Qualitatively, the experimental data are quite consistent with the local density approximation (LDA) calculations presented in Fig. \ref{fig1_map_GMX}(b).

Figures \ref{fig1_map_GMX}(e) and \ref{fig1_map_GMX}(f) show the 2.8 eV wide 2D curvature \cite{Zhang_RSI82} intensity plots along the $\Gamma$-M-X-$\Gamma$ momentum path at $k_z\sim 0$ and the Z-A-R-Z momentum path at $k_z\sim\pi$, respectively. For comparison, we superimpose the LDA bands on top of the experimental data. The experimental band dispersions show an excellent agreement with the LDA calculations after an overall renormalization by a factor of 1.1. 

For a closer look at the low-energy states, we present in Fig. \ref{fig2_Ascuts} ARPES intensity cuts along some high-symmetry lines, along with the corresponding intensity plots of curvature and either the related energy dispersion curve (EDC) or momentum distribution curve (MDC) plots. In particular, the top row shows clearly that while only two bands (labeled as $\alpha$ and $\beta$) cross $E_F$ along $\Gamma$-M ($k_z=0$), a third band ($\gamma$) emerges at $E_F$ along Z-A ($k_z=\pi$), which is consistent with our FS mappings. The $\eta$ and $\delta$ bands together form a hole pocket around the X point.



As compared with the Fe-based superconductors, the LDA calculations match the experimental results very well and the band renormalization is very small for SrNi$_2$As$_2$. The small renormalization factor (1.1) indicates that the electronic correlations in SrNi$_2$As$_2$ are much weaker than that in the Fe-based superconductors, in which the overall bandwidth is typically renormalized by a factor of 2 $\sim$ 5 \cite{0034-4885-74-12-124512,AmbroiseCRPhys2016}. In addition, we do not observe in SrNi$_2$As$_2$ any indication of orbital-selectivity, which contrasts with the Fe-pnictide superconductors, for which the bands with $d_{xy}$ character are renormalized twice as much as the other bands \cite{PhysRevX.3.011006}. The weak electronic correlations in SrNi$_2$As$_2$ go hand in hand with small local moments and absence of magnetic ordering. In an itinerant picture, the absence of good FS nesting conditions prevents the formation of a spin-density wave. 

\begin{figure}[!t]
\begin{center}
\includegraphics[width=2.7in]{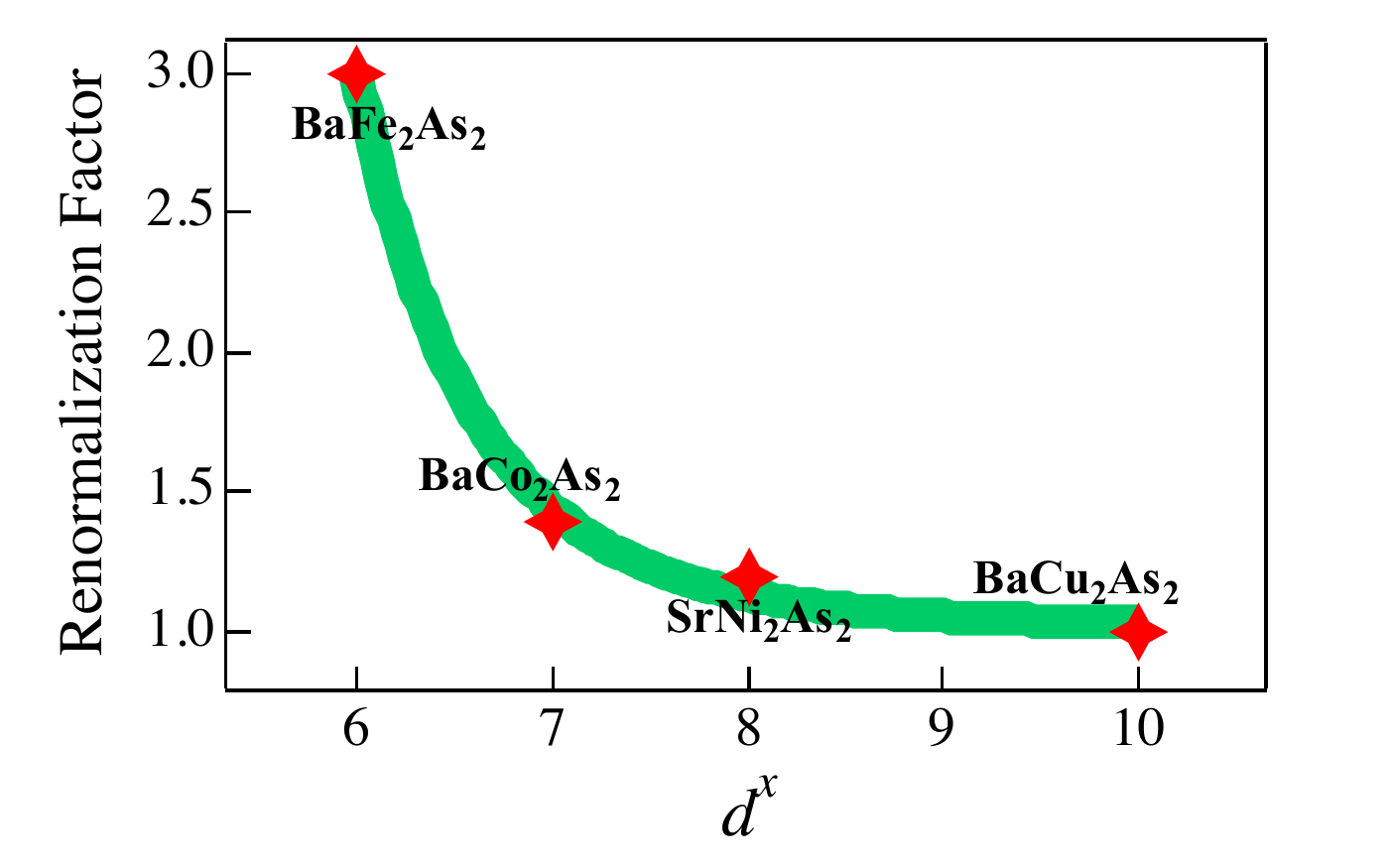}
\end{center}
\caption{\label{fig3_correlation}(Color online) Relationship between the renormalization factor and the number of $3d$ electrons for Fe- \cite{PhysRevLett.104.137001}, Co- \cite{PhysRevX.3.011006}, Ni- and Cu-based \cite{Shangfei_Wu_PRB91} 122-structure materials.}
\end{figure} 

In Fig. \ref{fig3_correlation} we compare the renormalization factor of different transition metal pnictides with the 122 structure. A clear monotonic but nonlinear decrease in the overall renormalization factor is observed as the filling of the $3d$ shell increases from $3d^6$ in BaFe$_2$As$_2$ \cite{PhysRevLett.104.137001} to $3d^{10}$ in BaCu$_2$As$_2$ \cite{Shangfei_Wu_PRB91}. This is consistent with the scenario that the electronic correlations are tuned by the electron filling of the $3d$ shell \cite{Razzoli_PRB91}. Although the Hund's coupling is vital in Fe compounds, we find in our calculations that its role for the electronic correlations are limited in the Co- and Ni- compounds. The sudden drop of electronic correlation strength from Fe-compounds to Co-compounds followed by a much smaller slope is mainly due to $3d$ electron filling.

To address the effect of the Hund's coupling and the influence of the filling of the Ni-$3d$ shell in SrNi$_2$As$_2$, we have performed dynamical mean-field theory (DMFT) calculations of the self-energy (for a review see \cite{Biermann_review}) at different physical and artificial fillings. The effective Hubbard interactions have been calculated within the constrained random phase approximation (cRPA) \cite{AryasetiawanPRB70} in the implementation of Ref. \cite{VaugierPRB86}. The construction of the Hamiltonian corresponds to what is commonly called a ``$d$-$dp$ Hamiltonian" in the literature \cite{Aichhorn_PRB80,Miyake_JPSJ77}, that is, Ni $d$- and As $p$-derived bands are included in the non-interacting part of the Hamiltonian, and Hubbard interactions are added for the Ni $d$ states. The values obtained for the Slater integrals are: $F_0=3.55$ eV, $F_2=8.08$ eV, $F_4=5.78$ eV. This corresponds to a Hubbard energy $U=3.55$ eV and a Hund's coupling $J=0.99$ eV, which is notably stronger than in BaFe$_2$As$_2$ or BaCo$_2$As$_2$ \cite{van_RoekeghemPRL113, PhysRevX.3.011006}. Double counting of correlations is avoided using the correction of the ``Around-Mean-Field" form \cite{AnisimovPRB44}. 

\begin{figure}[!t]
\begin{center}
\includegraphics[width=3.5in]{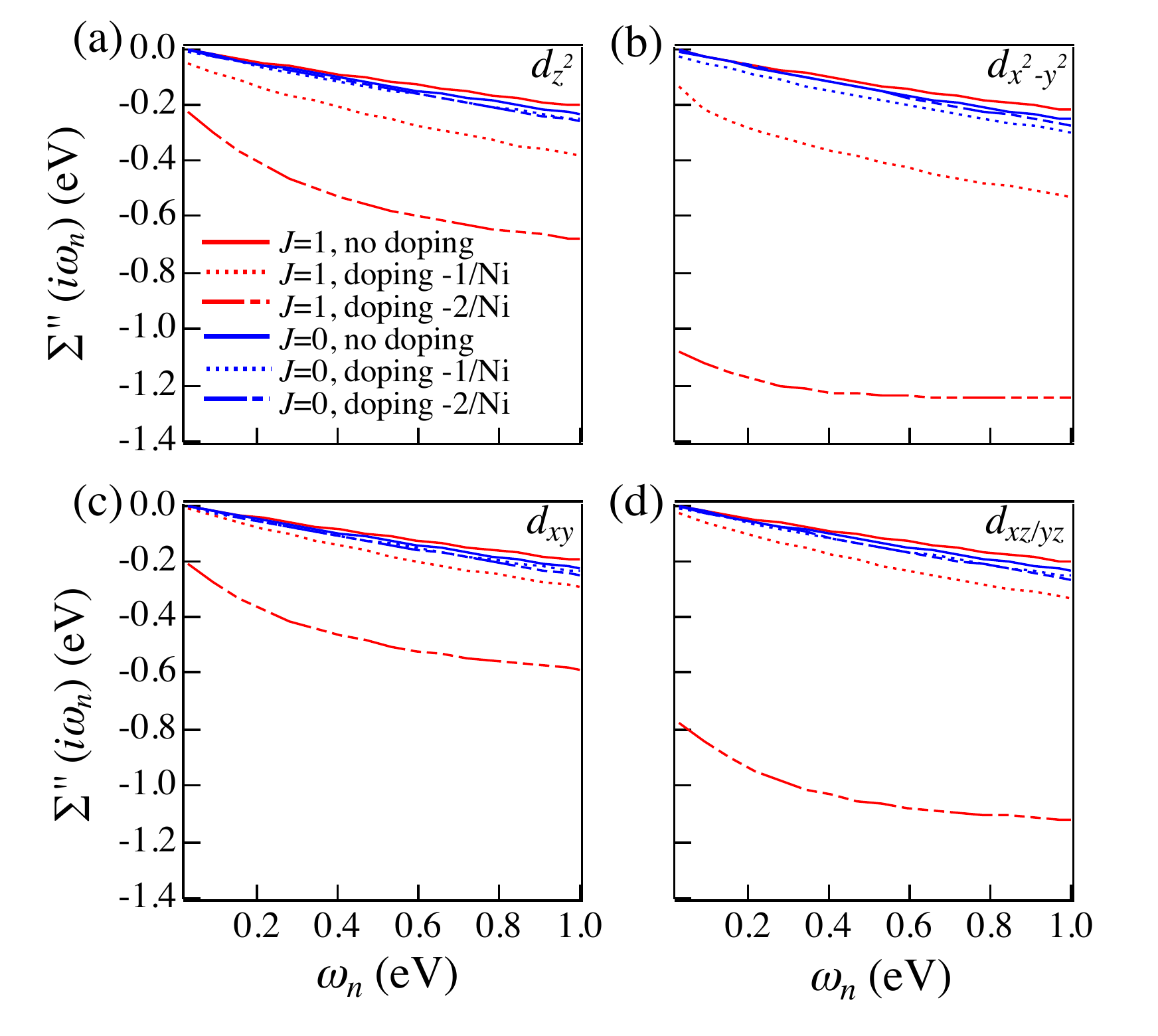}
\end{center}
\caption{\label{fig4_SelfE}(Color online) Imaginary part of the self-energy for orbitals (a) \textit{d$_{z^2}$}, (b) \textit{d$_{x^2-y^2}$}, (c) \textit{d$_{xy}$}, (d) \textit{d$_{xz/yz}$} as a function of the Matsubara frequency for SrNi$_2$As$_2$ and hypothetical compound consisting of SrNi$_2$As$_2$ with one and two electrons removed with respect to the filling of the physical SrNi$_2$As$_2$ compound.}
\end{figure}

\begin{figure}[!t]
\begin{center}
\includegraphics[width=2.7in]{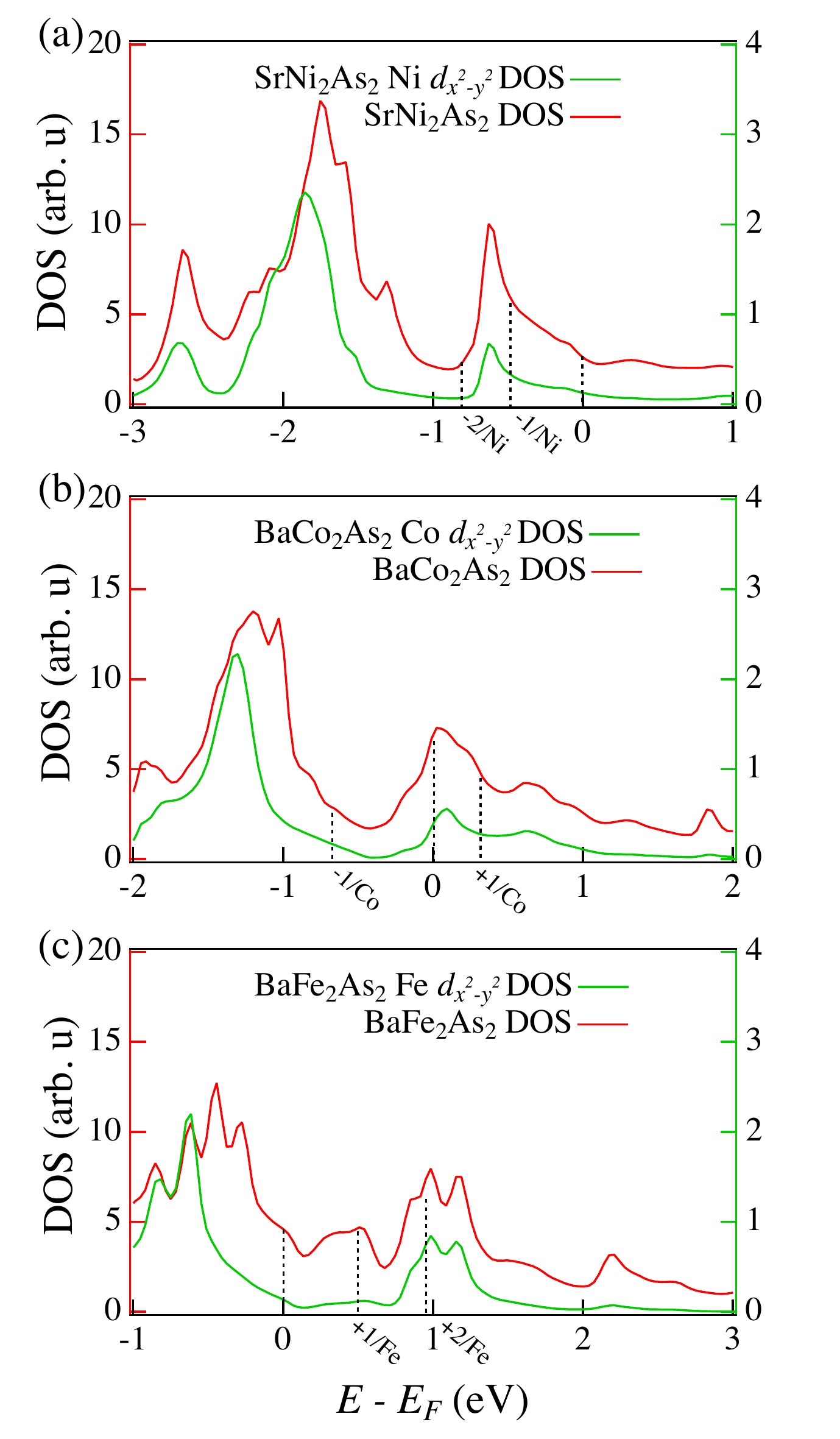}
\end{center}
\caption{\label{DOS}(Color online) Total and \textit{d$_{x^2-y^2}$} density of states for (a) SrNi$_2$As$_2$, (b) BaCo$_2$As$_2$, (c) BaFe$_2$As$_2$. The dashed lines correspond to doping levels 0, -1, -2 electron(s) per Ni for SrNi$_2$As$_2$, +1, 0, -1 electron(s) per Co for BaCo$_2$As$_2$ and +2, +1, 0 electron(s) per Fe for BaFe$_2$As$_2$, respectively. }
\end{figure}

We perform calculations for SrNi$_2$As$_2$ and for artificial compounds where we removed 1 and 2 electrons with respect to the filling of the physical SrNi$_2$As$_2$ compound. This is done by determining the chemical potential corresponding to these respective electron counts. In Fig. \ref{fig4_SelfE}, we show the imaginary part of the self-energy $\Sigma$ as a function of (imaginary) Matsubara frequencies for all three situations. We also compare this result with analogous calculations in which we put  the Hund's coupling to zero (\textit{i.e.} $F_2=F_4=0$ in terms of the Slater integrals).  In a Fermi liquid, the imaginary part of $\Sigma$ on the Matsubara axis vanishes linearly, and the slope is related to the quasi-particle residue $Z$ as 

\begin{eqnarray}
Z = \frac{1}{1 - \frac{\partial \Sigma(i \omega)}{\partial i \omega}}
\end{eqnarray}

Within DMFT, $\Sigma$ is purely local and $Z$ directly equals the inverse effective mass enhancement. For SrNi$_2$As$_2$, we obtain an effective mass of m*/m=1.2 $\sim$ 1.3 in good agreement with the experiment. The orbital dependence of the effective mass enhancement is negligible in this case. Interestingly, the effect of the electronic correlations is found to be even weaker than in the case of vanishing Hund's coupling $J=0$, in contrast to what was found in BaFe$_2$As$_2$ or BaCo$_2$As$_2$ \cite{WernerNPhys8, PhysRevX.3.011006, van_RoekeghemPRL113}. Such a behavior is consistent with the scenario of the influence of Hund's coupling developed in the literature \cite{WernerPRL101, WernerNPhys8, HauleNJP11, de_MediciPRL107}. The key parameter is the large filling of the Ni-$3d$ shell. Indeed, the constructed Wannier orbitals contain 8.8 electrons due to the Ni-As hybridization that increases the number of electrons in the Ni-$3d$ shell with respect to the nominal electron count of 8 electrons per Ni. When this number of electrons is reduced by 1 per Ni atom, the effect of correlations is slightly enhanced by the Hund's coupling and a small orbital dependency is retrieved. 

In Fig. \ref{DOS} we note in particular the more strongly correlated behavior of the $d_{x^2-y^2}$-derived band, which has a large density close to $E_F$. This effect becomes huge when 2 electrons per Ni atom are removed. It is even more pronounced than in BaFe$_2$As$_2$ due to the larger value of the Hund's coupling. Interestingly, the most correlated orbital in this case is still $d_{x^2-y^2}$, in contrast to BaFe$_2$As$_2$ where at this filling the  $d_{xy}$ orbital is the most strongly correlated. We have traced back this difference to a more pronounced 3D character in SrNi$_2$As$_2$ caused by the reduction of the $c/a$ ratio with respect to the ones of BaFe$_2$As$_2$ and BaCo$_2$As$_2$. Indeed, the larger dispersion of out-of-plane bands, in particular in the  $k_z=\pi$ plane, leaves the flat $d_{x^2-y^2}$ band at $E_F$ when 2 electrons per Ni are removed in the Ni compound (the band around \textit{E$_B$} = -0.808 eV in Fig. \ref{fig1_map_GMX} (f)), while the density-of-states of this orbital presents a pseudo-gap near $E_F$ in the Fe compound \cite{Takashi_JPSJ}, \textit{i. e.} a strong suppression of density-of-states, as shown by the green curve in Fig. \ref{DOS}(c). The presence of the flat band at $E_F$ is strongly reminiscent to the case of undoped BaCo$_2$As$_2$ \cite{PhysRevX.3.011006,Dhaka_PRB87}. In the latter compound, the physics is very strongly dependent on the precise energetic position of this flat band, which determines its paramagnetic behavior (in contrast to ferromagnetic CaCo$_2$As$_2$ \cite{ChengPRB85}). Yet, the overall degree of correlations in BaCo$_2$As$_2$ is weak. This suggests the global filling to be the key tuning parameter for the correlation strength. The details of the density-of-states, including van Hove singularities close to $E_F$, then intervene to tune the orbital-dependence of the correlations. Finally, we note again that the effect of correlations is weak and largely independent of the filling when no Hund's coupling is present. This demonstrates once more the importance of this quantity to tune the electronic correlations in transition-metal pnictides, from stronger -- as in Fe-pnictides -- to weaker correlations -- as in SrNi$_2$As$_2$.

\section{\label{Summary}Summary}

In summary, we have studied the electronic structures of the Ni-based superconductors SrNi$_2$As$_2$. One additional hole Fermi surface around BZ center Z point is observed compared to $\Gamma$ point, which indicates a Strong three-dimensionality. Our results are in good agreement with LDA calculations, albeit for a small overall, orbital insensitive renormalization factor of 1.1, suggesting that the electronic correlations in these compounds are weak compared to the Fe-based superconductors. The nonlinear reduction of the electronic correlation strength with the increasing filling of the $3d$ shell for the Fe-, Co- and Ni-based compounds confirms that $3d$ electron filling plays an important role in tuning the electronic correlations, which may be related to higher $T_c$'s in the Fe-based materials. Although we observe both hole and electron FS pockets separated by the same wave vector as in the Fe-based superconductors, the FS nesting is quite poor, which is possibly a reason for the absence of long-range magnetic order in SrNi$_2$As$_2$, along with the negligible local moments.
  
 \section{Acknowledgement}
 
We acknowledge X. X. Wu for useful discussions. This work was supported by grants from MOST (2011CBA001000, 2011CBA00102, 2012CB821403 and 2013CB921703 and 2015CB921301) and NSFC (11004232, 11034011/A0402, 11234014 and 11274362) from China,
as well as by a Consolidator Grant (project number 617196) of the European
Research Council.

\bibliography{Manuscript_Zeng}

\end{document}